\documentclass{icrc29}
\usepackage{graphicx,amssymb,amsmath,times}
\begin{document}
\title{Coverage and large scale anisotropies estimation methods for the Pierre Auger Observatory}
\author[The Pierre Auger Collaboration]{The Pierre Auger Collaboration\\Av. San Mart\'{\i}n Norte 304, (5613) Malarg\"ue, Argentina }
\presenter{Presenter: J.-Ch. Hamilton (hamilton@in2p3.fr), fra-hamilton-JC-abs1-he14-oral}

\maketitle

\begin{abstract}
When searching for anisotropies in the arrival directions of Ultra High Energy Cosmic Rays, one must estimate the number of events expected in each direction of the sky in the case of a perfect isotropy. We present in this article a new method, developed for the Auger Observatory, based on a smooth estimate of the zenith angle distribution obtained from the data itself (which is essentially unchanged in the case of the presence of a large scale anisotropy pattern). We also study the sensitivity of several methods to detect large-scale anisotropies in the cosmic ray arrival direction distribution : Rayleigh analysis, dipole fitting and angular power spectrum estimation.
\end{abstract}

\section{Introduction}

The large scale distribution of the arrival directions of UHECR is among the observables that might help to solve the UHECR puzzle in the next few years thanks to the large statistics collected by the Auger Observatory~\cite{auger}. At low energies ($<$ 1 EeV), the arrival directions are expected to be isotropized by the Galactic magnetic fields. At higher energies ($>$ 10 EeV), sources may possibly show up as excess of showers from one direction and therefore be detected. It is also between these energies that the cosmic rays are expected to change from a galactic to an extragalactic origin so that their large scale angular distribution might change significantly, giving precious hints on the origin and nature of these particles and the magnetic fields that modify their trajectories. The AGASA claim for large scale anisotropy~\cite{agasa} is another motivation for large scale structure search. Such a search relies heavily on the estimation of the background number of cosmic rays expected from each sky direction in order to disentangle acceptance effects from real anisotropies on the sky. We shall call this the {\em coverage map} from now on. In the present article, we propose a semi-analytical estimation of the coverage map based on a maximal use of the symmetries of the Pierre Auger surface detector (almost uniform acceptance in sidereal time and azimuth) and on the fact that the zenith angle distribution of the events is almost unaffected by the possible presence of sky anisotropy. We also present techniques that the Auger experiment intends to use for large scale anisotropy search: dipole fitting, first harmonic analysis and angular power spectrum estimation. 

\section{Semi-analytical coverage map estimation}
\begin{figure}[!t]
\resizebox{\hsize}{!}{\centering{\includegraphics{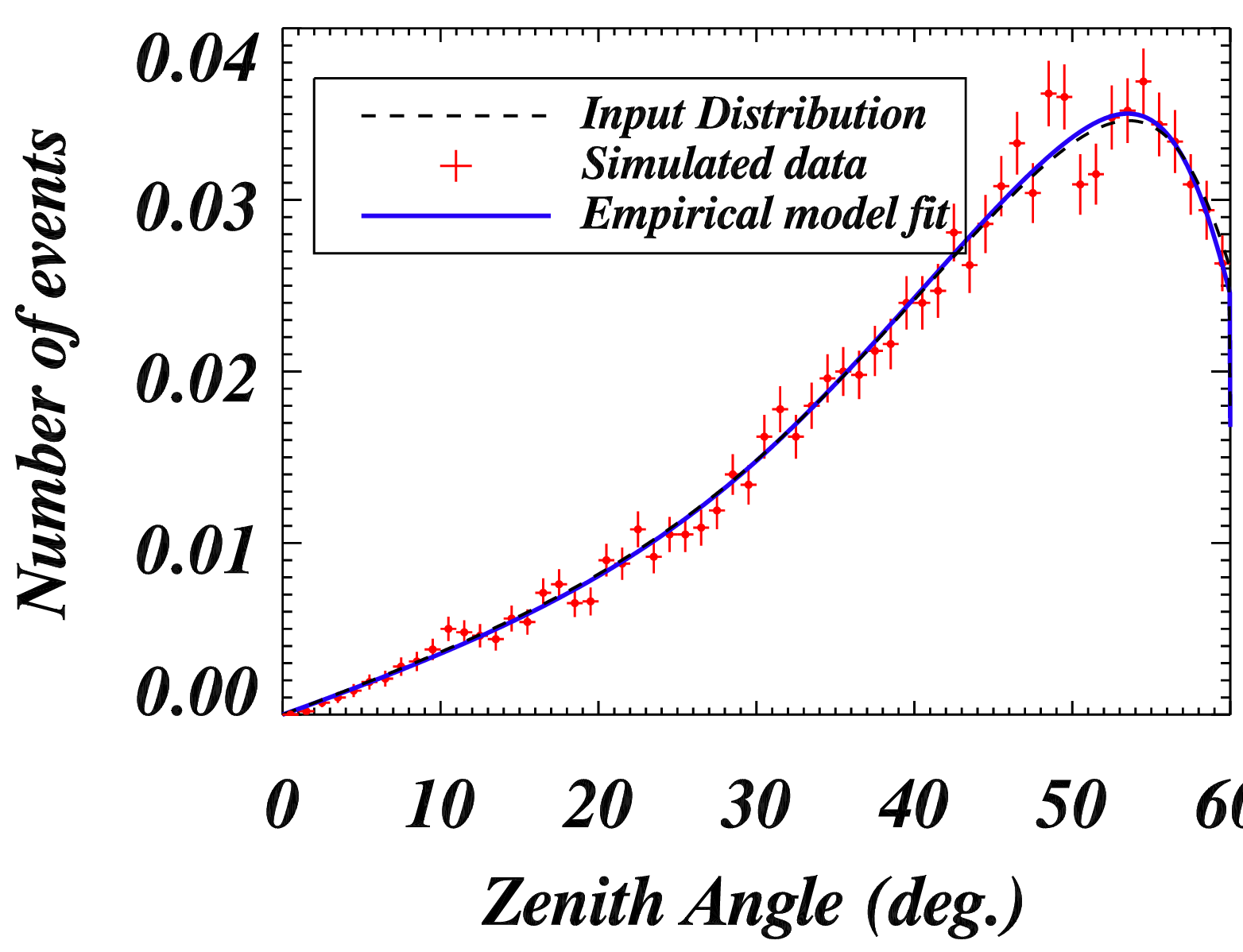}\includegraphics{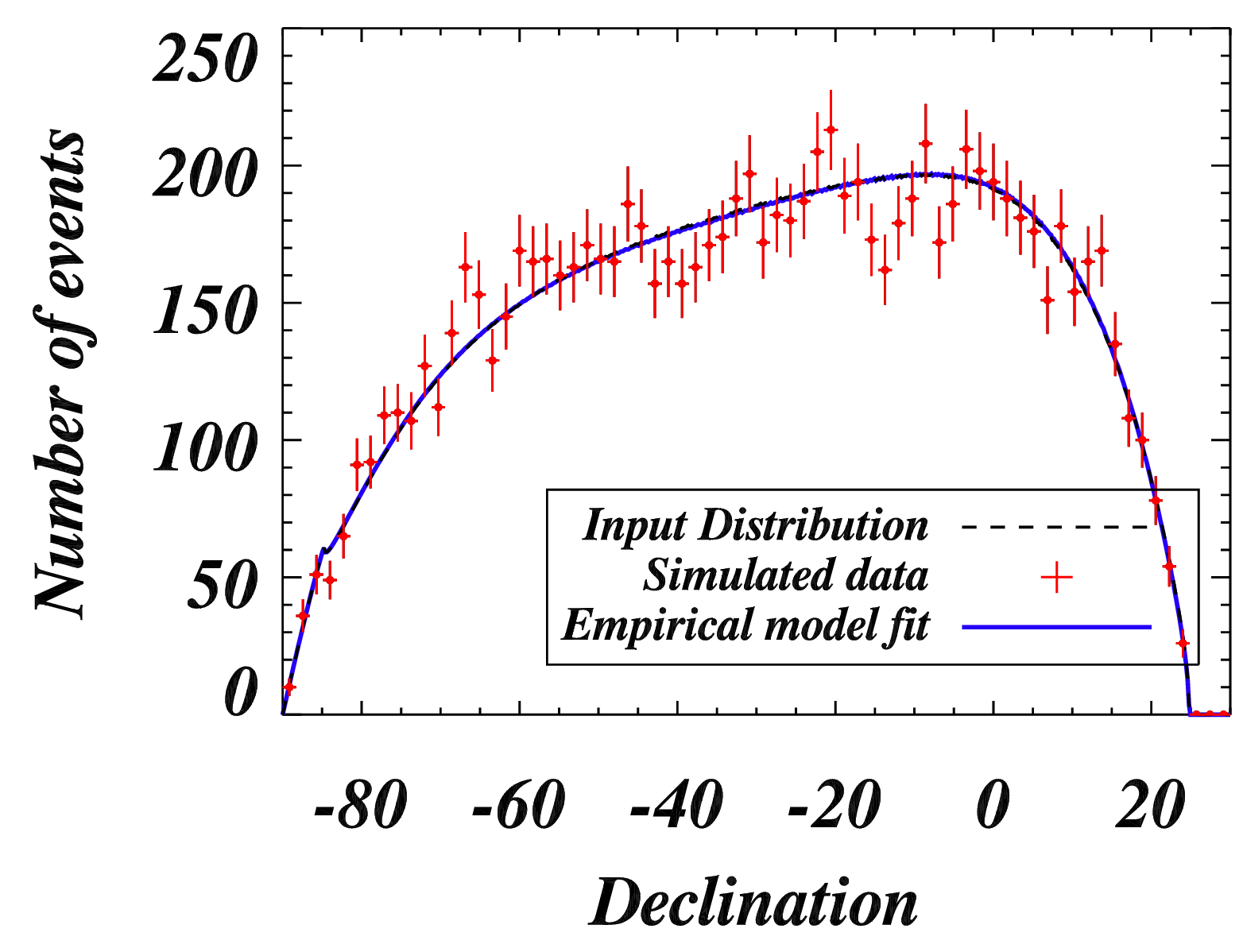}\includegraphics[angle=90,scale=0.8]{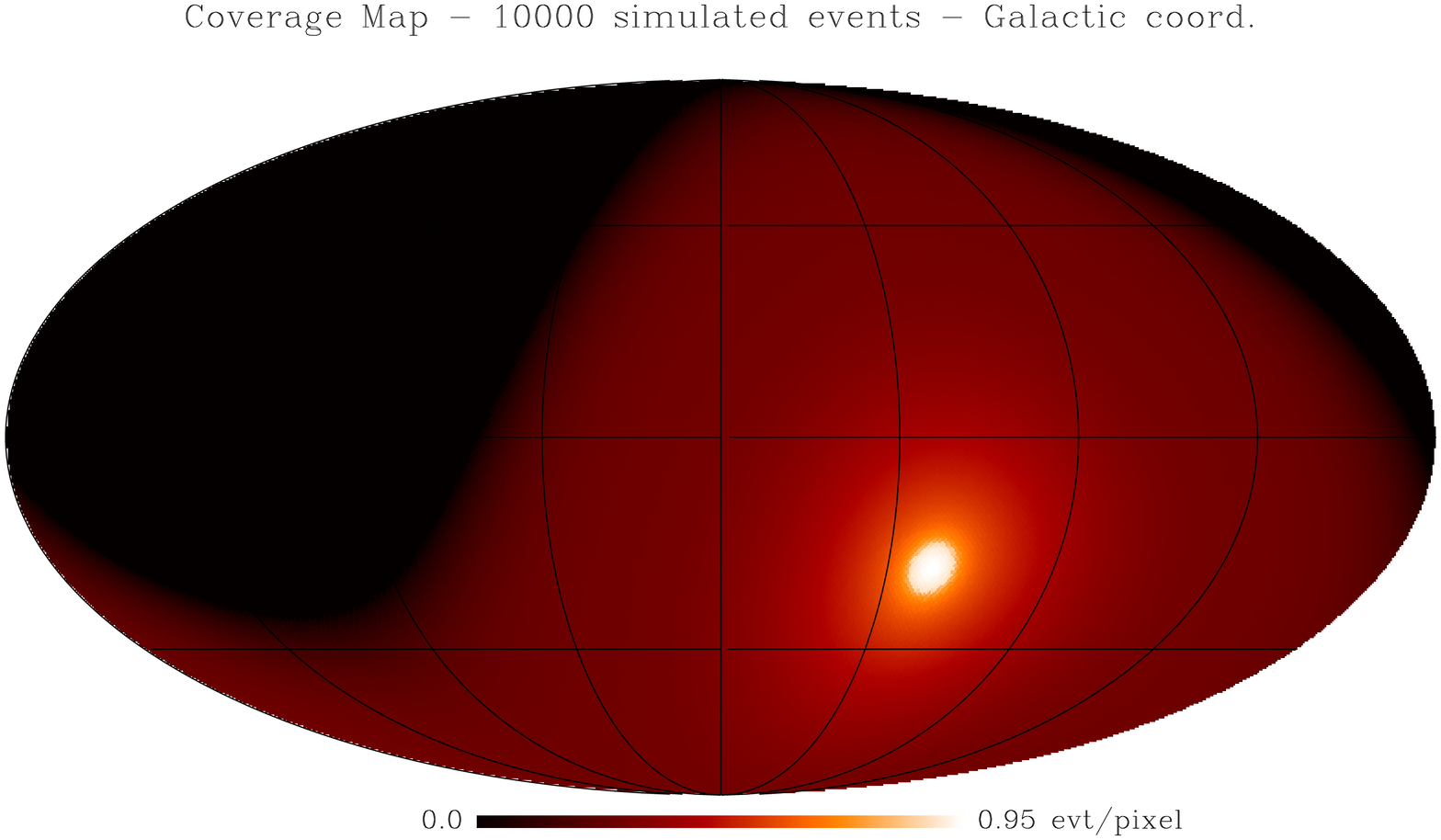}}}
\caption{Results obtained with an isotropic simulation with 10000 events: Zenith angle distribution (left) and empirical model fit, resulting declination coverage along with the data (center) and final coverage map (right) in Galactic coordinates using a Mollweide projection and the Healpix~\cite{healpix} pixellisation. The true zenith angle and declination distributions are shown in black dashed lines and can hardly be distinguished from the fit.}
\label{fig1}
\end{figure}

The Auger surface detector is designed in such a way that its acceptance is almost independent of both sidereal time and azimuth. The zenith angle of an event is related to the equatorial coordinates through $\cos\theta(\tau) = \sin\delta \sin l +\cos \delta \cos l \cos(\alpha-A(\tau))$ where $l$ is the latitude of the experiment ($-35.2^\circ$ for the Auger Southern site). $A(\tau)$ is the right ascension of the zenith at the observatory location at sidereal time $\tau$. Integrating the zenith angle acceptance $a[\theta]$ over a full sidereal day leads to a coverage map that is only a function of $\delta$: 
\begin{equation}\label{eq1}
W(\delta)=\int_0^\mathrm{24h} a\left[\theta(\alpha-A(\tau),\delta)\right]\mathrm{d}\tau
\end{equation}
The acceptance per unit solid angle is proportional to the geometrical factor $\cos\theta$ but in addition large zenith angle showers, especially at low energies, are known to be attenuated by the larger atmospheric depth they have traveled through and this induces a cutoff at large angles. In order to account for this as well as for more complex acceptance effects, we do an empirical fit of the zenith angle distribution by the geometrical acceptance multiplied by a Fermi-Dirac function and $n$-parameters splines (or equivalently, polynomials). The fit to real data is in general satisfactory with $n = 3$. The sky coverage is then obtained by numerical integration of Eq.~\ref{eq1}. We show in Fig.~\ref{fig1} the fit zenith angle distribution on simulated data, the resulting declination distribution and coverage map. This method relies on the assumption that the acceptance is independent of sidereal time and azimuth. It can be relaxed by replacing the zenith angle dependent acceptance by a more complex function depending also on time for instance. Such a dependence exists with the Auger surface detector as the array grows with time, it could also account for daily and seasonal modulations due to temperature and pressure effects on the trigger rate~\cite{parizot}. Now each direction of the sky $(\alpha,\delta)$ corresponds at UTC $t$ to direction $(\theta(t), \phi(t))$. The coverage map is then obtained by integrating numerically the global acceptance $a_{\mathrm{tot}}(t) = a \left[ \theta(t), \phi(t) \right]$ over the whole data taking period: 
\begin{equation}\label{eq2}
W(\alpha,\delta)=\int_{t_{min}}^{t_{max}}a_\mathrm{tot}(t)\mathrm{d}t
\end{equation}
A satisfactory model for the acceptance as a function of time is not easy to obtain, but we assume here that it is known. We see that the simple expression in Eq.~\ref{eq1} is obtained with Eq.~\ref{eq2} with a constant time and azimuth acceptance and therefore replacing the universal time integration by a sidereal time integration over 24 hours. 

The presence of anisotropies on the sky could lead to a biased coverage map because of the induced modification of the zenith angle distribution, resulting in an underestimation of the real sky anisotropy. The effect has however been shown to be negligible because the anisotropy is largely averaged in zenith angle space (see Fig.~\ref{eq2}-left for a 50\% dipole oriented towards the South Equatorial Pole). The distortion is maximal for an anisotropy oriented towards the equatorial poles ({\em ie} large zenith angle in local coordinates). This was tested (for a time independent acceptance) using simulated data with fake dipoles ($5\%$ amplitude) oriented in various directions. The impact on the coverage map is always smaller than $5\%$ with our semi-analytical (hereafter SA) method. We compare our results with those of the scrambling method~\cite{clay} which consists in averaging a large number of fake data sample by exchanging sidereal times and azimuth of the events. Fig.~\ref{eq2} shows the relative coverage bias for the SA and scrambling methods for a $5\%$ dipole towards the South Pole. Two flavors of the scrambling method were tested: the usual one (described above and quoted as 2D) and a version (quoted as 1D) where azimuths are drawn uniformly instead of being scrambled so as to implement the uniform azimuth acceptance assumed in the SA method. The results are always significantly better with our method. We therefore conclude that if we are able to model properly the possible time dependence of the experiment, it is preferable to construct the coverage map with the SA method.
\begin{figure}[!t]
\resizebox{\hsize}{!}{\centering{\includegraphics{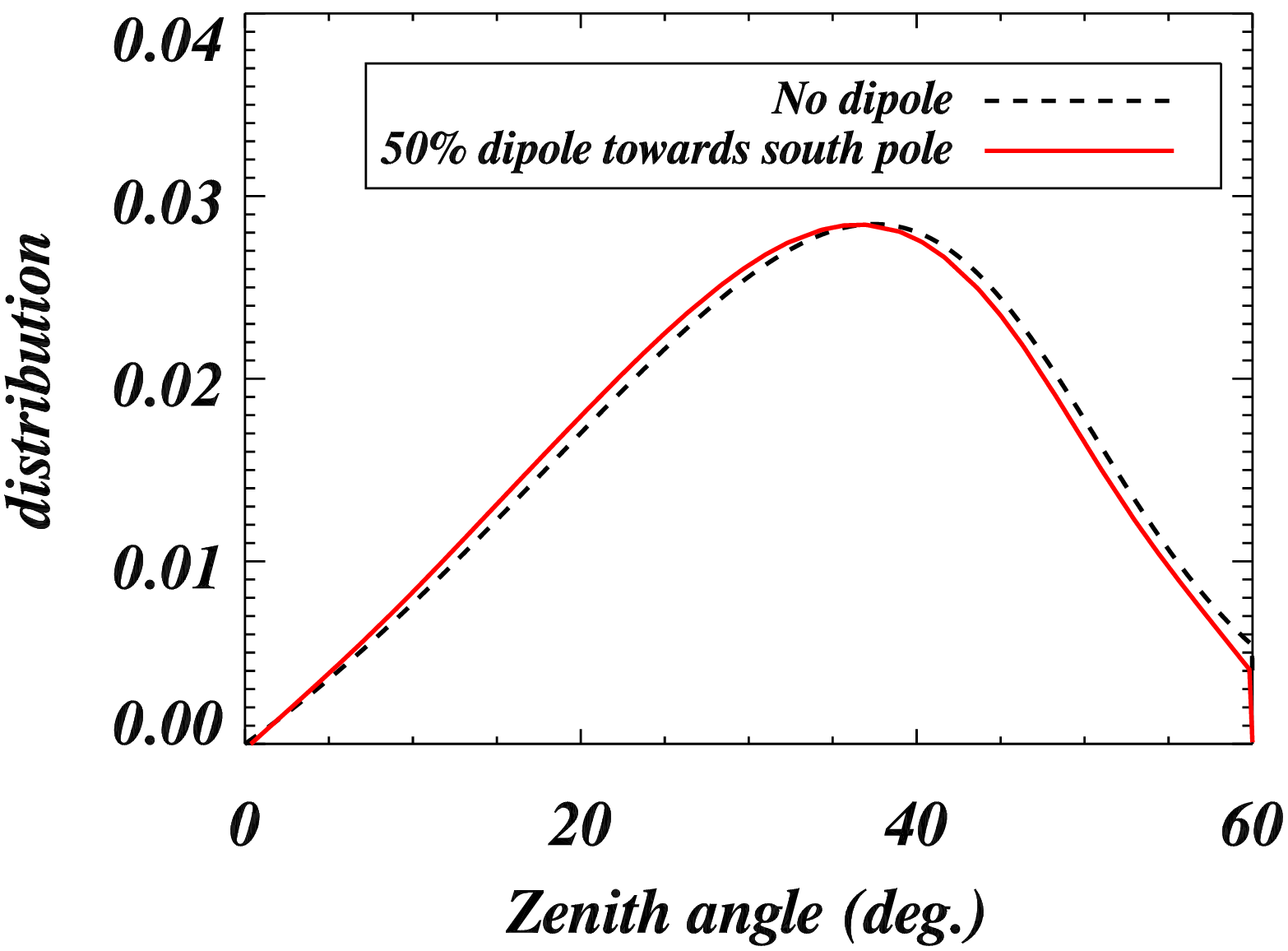} \includegraphics{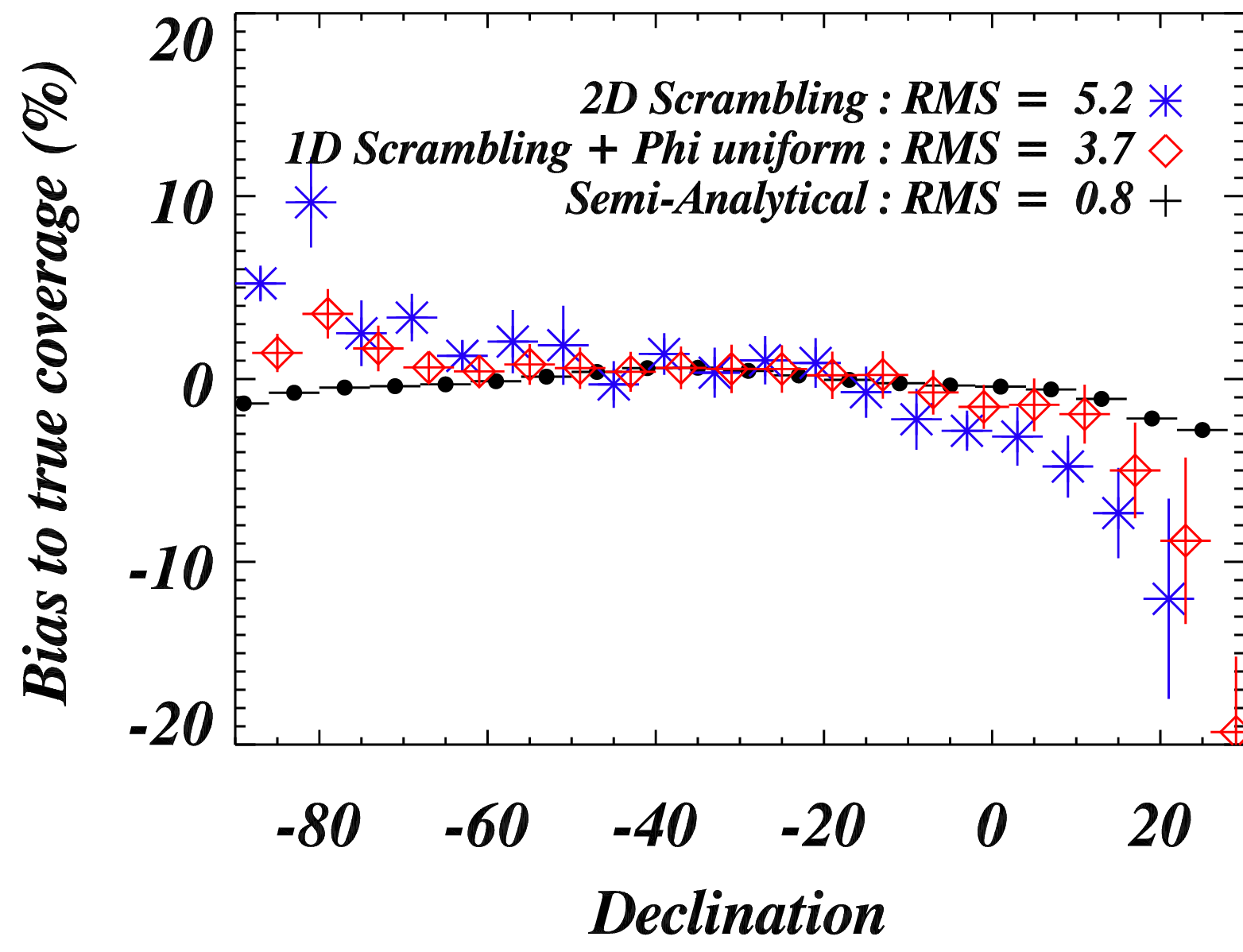}\includegraphics{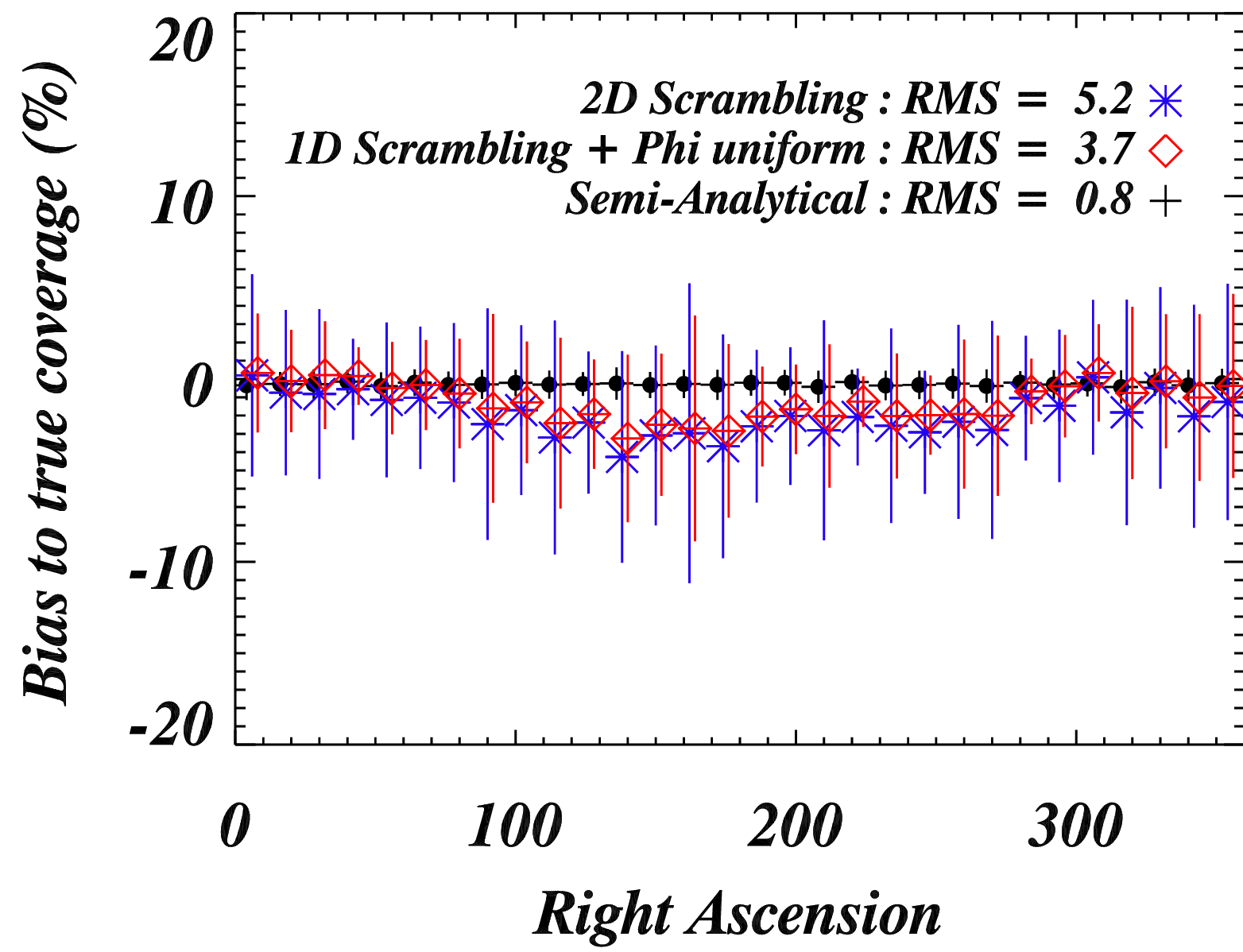}}}
\caption{Effect of a $50\%$ dipole towards the South pole on the zenith angle distribution (left). Relative bias to true coverage map for the SA method (black crosses), the 2D scrambling (blue stars) and 1D scrambling with uniform phi drawing (red diamonds) as a function of declination (center) and right ascension (right). The error bars reflect the RMS dispersion in each bin. We used a $5\%$ dipole pointing to the South Pole. }
\label{fig2}
\end{figure}

\section{Methods for large scale anisotropy search}
A common method widely used in earlier cosmic rays analyses is the Rayleigh analysis. It consists in computing the Fourier first harmonic amplitude of the right ascension distribution in the dataset. This exploits the stable running for a long period that should lead to an approximately constant sidereal time exposure. It does not require the knowledge of the exposure declination dependence since only information about the events right ascension is used. Consequently it only allows to recover a 2D projection of the original anisotropy. This can be improved by considering various declination bins but it is not a true two-dimensional analysis. Another approach is to fit or compute directly a dipole pattern on the data accounting for the coverage map. We have developed three different flavors all leading to comparable precisions. We show in Tab.~\ref{tab1} the resolutions obtained on a large number of simulations with a $5\%$ dipole towards $(\delta = -45^\circ, \alpha = 0^\circ)$ and 30000 events.
\begin{itemize}
\item {\bf Direct fit:} One directly fits the product of the coverage map with a general dipole (three degrees of freedom) superimposed on a uniform background. 
\item {\bf Dipole vector computation:} It is a generalization to a partial sky coverage of the one proposed in~\cite{sommers} and is described in detail in~\cite{sap}. The idea is to compute the average vector pointed to by the events weighted with the inverse coverage map: 
$\vec{D} = \frac{3}{N}\sum^n_{i=1} W^{-1}_i \vec{n}_i$, where $\vec{n}_i$ is the unit vector pointing towards event $i$, $W_i$ the coverage map in this direction, and the effective number of events is $\mathcal{N} = \sum^n_{i=1} W^{ -1}_i$. The dipole parameters (amplitude $a$ and direction) are obtained by simple algebraic identification of this vector with the one expected from a dipole modulated flux. With full sky coverage, $\vec{D}$ approximates the dipole vector itself, $a\times\vec{d}$, within $\sqrt{n}/a$. With partial sky coverage, it approximates a linear combination of the dipole parameters (requiring an additional straightforward inversion), with similar uncertainties depending on the fraction of the sky covered. 
\item {\bf $\chi^2$+Rayleigh:} Detailed in~\cite{bariloche}, it relies on the fact that the dipole component along the NS axis, $\alpha_z$, can be obtained by just fitting the declination distribution of the events with $\mathrm{d}N/\mathrm{d}\delta\propto W(\delta)(1+\alpha_z\cos\delta)$.  The value of $\alpha_z$ so obtained is unbiased, and combining it with the results of the Rayleigh analysis one recovers the three components of the dipole vector. Once the dipole orientation is known, one can fit the overall distribution to a dipolar one along this direction, and the value of the $\chi^2$ obtained will be indicative of the quality of the assumption that the anisotropy was of a dipolar type. This method can incorporate the effects of a right ascension modulation of the exposure and can also reconstruct a quadrupole component.
\end{itemize}
A natural extension of the above techniques to higher order harmonics is to expand the observed events distribution on the sky on the spherical harmonics basis. Similarly to CMB studies, one estimates the angular power spectrum $C_\ell$ (the variance of the coefficients of the expansion $a_{\ell m}$) of the data up to any $\ell$ multipole allowed by the resolution of the experiment. Under the assumption that the anisotropies are statistically homogeneous, the full sky power spectrum can be estimated even in case of partial sky coverage using the method proposed by~\cite{hivon} (already described in~\cite{peebles}) for CMB studies and reformulated for Cosmic Rays purposes in~\cite{deligny}. The angular power spectrum estimation is not a fit on the data but a harmonic space expansion. It is therefore not a problem to go for higher order multipoles. The price to pay is however that the orientation of the reconstructed patterns is lost as we only consider $C_\ell$ and not the full $a_{\ell m}$. 

\begin{table}[!t]
\centering
\begin{tabular}{c|cc|cc|cc}
&$<a>$ &$\sigma_a$ &$<\delta>$ &$\sigma_\delta$ &$<\alpha>$ &$\sigma_\alpha$ \\ 
\hline
Direct fit      &4.4\% &$\pm$ 1.3\% &-43.5$^\circ$ &$\pm$ 17.9$^\circ$ &3.0$^\circ$  &$\pm$ 19.6$^\circ$ \\
Dipole Vector   &5.8\% &$\pm$ 1.7\% &-44.6$^\circ$ &$\pm$ 19.1$^\circ$ &-0.3$^\circ$ &$\pm$ 21.6$^\circ$ \\
$\chi^2$+Rayleigh
                &5.4\% &$\pm$ 1.3\% &-44.5$^\circ$ &$\pm$ 16.1$^\circ$ &0.2$^\circ$ &$\pm$ 17.6$^\circ$
\end{tabular}
\caption{\small Dipole parameters (and dispersions) recovered with the various methods on a large number of simulations.}
\label{tab1}
\end{table}

\section{Conclusions}
We have presented a semi-analytical estimation of the expected number of background events for cosmic rays experiments. This is based on a smooth model of the acceptance as a function of zenith angle which is almost unchanged by the presence of anisotropies. The coverage map can account for complex acceptance effects such as azimuthal and sidereal time dependence. Our method is both more precise and less biased by possible true anisotropy on the sky than the usual scrambling method. We have also proposed various ways to search for such large scale anisotropies: first harmonic analysis, dipole orientation and amplitude determination and angular power spectrum determination. All of these methods (except the Rayleigh analysis) take profit of the accuracy of the coverage map estimate. This coverage map estimation technique can also be applied for small scale anisotropy searches~\cite{antoine,benoit}.

\end{document}